\documentclass{iopart}

\newcommand{\ai}{\'{\i}}
\renewcommand {\slash}[1]{#1 \!\!\! /}

\usepackage{epsfig}  

\begin{document}  

%\draft 

\title{Classical limit for the scattering of Dirac particles 
              in a magnetic field.}    
\author{G Murgu\ai{}a \footnote[1]{E-mail: murguia@fisica.unam.mx}
        and M Moreno \footnote[2]{E-mail: matias@fisica.unam.mx}
       }  

%ReVTeX 3.1 and IoP: 
\address{Instituto de F\ai{}sica, UNAM.   
         Apartado postal 20364, 01000, M\'exico, D.F. Mexico.}    

%ReVTeX 4:
%\affiliation{Instituto de F\ai{}sica, UNAM.
%         Apartado postal 20364, 01000, M\'exico, D.F. Mexico.}  

%\date{\today}  
  
%ReVTeX 3.1: 
%\maketitle  
 
%IoP:
%\ead{\mailto{murguia@fisica.unam.mx}
%     \mailto{matias@fisica.unam.mx}}

%%% Abstract %%%   
\begin{abstract}  
    
  We present a relativistic quantum calculation at first order in
  perturbation theory of the differential cross section for a Dirac
  particle scattered by a solenoidal magnetic field. The resulting
  cross section is symmetric in the scattering angle as those obtained
  by Aharonov and Bohm~(AB) in the string limit and by Landau and
  Lifshitz~(LL) for the non relativistic case. We show that taking
  $pr_0\left|{\sin{(\theta/2)}}\right|/\hbar\ll 1$ in our expression
  of the differential cross section it reduces to the one reported by
  AB, and if additionally we assume $\theta \ll 1$ our result becomes
  the one obtained by LL.  However, these limits are explicitly
  singular in $\hbar$ as opposed to our initial result. We analyze the
  singular behavior in $\hbar$ and show that the perturbative Planck's
  limit ($\hbar \rightarrow 0$) is consistent, contrarily to those of
  the AB and LL expressions. We also discuss the scattering in a
  uniform and constant magnetic field, which resembles some features of
  QCD.
  
\end{abstract}  

%IoP:
%\submitto{\JPA}


%ReVTeX 4 and IoP:
\maketitle  
  
%%% Section: Introduction %%%   
\section{Introduction.}   
  
We know that in the classical scattering of charged particles by
magnetic fields the scattered particles describe circular trajectories
with fixed radii, so they have a preferential direction of motion. In
this work, the relativistic quantum version of the problem is studied
in the lowest order of perturbation theory and a symmetric behavior in
the scattering angle is found for uniform and solenoidal magnetic
fields. We focus on the study of some interesting limit cases of
the differential cross section and compare them with previous non
relativistic results reported by Aharonov and Bohm (AB)~\cite{AB} and
Landau and Lifshitz (LL)~\cite{LL}.

Although the scattering of charged particles by a solenoidal magnetic
field, Aharonov-Bohm (AB) effect has been studied perturbatively
before by other authors~\cite{Boz}, in this work we are interested in the
classical limit of the differential cross section.

As it is known, the AB effect~\cite{AB} is considered one of the most
important confirmed~\cite{Chambers-Tonomura} predictions of the
quantum mechanics because it shows that the vector potential has a
physical significance and can be viewed more than a mathemathical
convenience. The interest in this effect has been increased
recently.
%~\cite{Varios}%
Both because of basic reasons that have
changed the understanding of gauge fileds and forces in nature and
also because it has a lot of connections with new physics, like the
quantum Hall effect~\cite{QHE}, mesoscopic physics~\cite{MesoscopicP}
and physics of anyons~\cite{Anyons}. 

In the last two decades, several treatments to the AB scattering
problem have appeared.
First, the magnetic phase factor in the non
relativistic case was treated directly ({\it e.g.}~\cite{Berry}).
Works that consider spin have also been carried out and specially the
behavior of the wave function is studied taking into account a delta
function like potential at the origin~\cite{Hagen_90}. Although
originally AB solved the problem exactly, the perturbative analysis has
played an important role~\cite{Hagen_95,Gomes,Boz}, giving rise to
discussions about the form in which the incident wave must be
treated~\cite{Sakoda}. Also, various QED processes of scalar and Dirac
particles in the AB potential have been carried out, and special
interest has been devoted to polarization properties in bremsstrahlung
and synchrotron radiation~\cite{Audretsch,Bagrov}.

Our interest in this problem stems from the fact that an electron
in a uniform infinite magnetic field is trapped in two dimensions in a
potential of the harmonic oscillator form. Thus it is a confined
point-like fermion and therefore resembles the dynamic confinement
produced in Quantum Chromodynamics (QCD) for quarks in three
dimensions. In order to keep our analogy as close as possible to the
parton model ideas our computation is done in a perturbative fashion.
In this model, the use of free particle asymptotic states is very
common, nevertheless with a simple model calculation in Quantum
Electrodynamics (QED) we show that this procedure could be misleading
at least in lowest order.

In this work we focus our analysis in the classical limit. First we
are puzzled by the symmetric results of the differential cross section
with respect to the scattering angle $\theta$, in contradistinction to
the classical scattering which favor an asymmetrical result. Second,
the classical limit of field theory is a long unsolved problem, it is
therefore tempting to understand the above mentioned quantum
(symmetric) vs classical (asymmetric) results in order to clarify the
classical limit: Are we in the presence of a process which is purely
quantum in nature as LL suggest?

%%% Section: Non-relativistic case %%%  
\section{Previous results in the non relativistic case.}  
  
Let us recall two landmark results of the non relativistic case for  
the differential cross section of the scattering of electrons by  
solenoidal magnetic fields.  
Chronologically, the first result was presented by Aharonov and Bohm  
\cite{AB}. They obtain the exact solution for the scattering problem  
when the radius of the solenoid is very small for a constant magnetic  
flux, in fact they consider only a quanta of magnetic flux ($\Phi =  
2\pi\hbar c/e \sim 4 \times 10^{-7}$ gauss cm$^2$). Their result 
is\footnote[1]{In the paper of Aharonov and Bohm the cross section appears
to be inversely proportional to $\cos^2{\theta/2}$, but the reference
frame they use is such that $\theta$ is traslated by $\pi$. Also the
factor $\hbar/p$ is not explicitly shown due to a change of variable
($r' = kr$).}:
\begin{equation}  
\left.{\frac{d\sigma}{d\theta}}\right|_{AB}=  
         \hbar \frac{\sin^2{\left({\frac{e\Phi}{2\hbar c}}\right)}}  
         {2 \pi p \sin^2{\frac{\theta}{2}}}.  
\label{secc.dif.AB}  
\end{equation}  
Independently, Landau and Lifshitz~\cite{LL} study the  
same scattering problem with the use of the eikonal  
approximation. Including only the contribution of the vector potential  
from the exterior of the solenoid, they obtain  
precisely the same as Aharonov and Bohm. Notice that this cross  
section is symmetric in $\theta$.  
  
Landau and Lifshitz compute the differential cross section for small
scattering angles in the case of a small magnetic flux, $e\Phi/2\hbar
c \ll 1$, where perturbation theory is applicable, then
$\sin{(e\Phi/2\hbar c)} \approx e\Phi/2\hbar c$, and the resulting
cross section develops a singular behavior in $\hbar$:
\begin{equation}  
\left.{\frac{d\sigma}{d\theta}}\right|_{LL; \theta \ll 1}=  
       \frac{e^2\Phi^2}{2\pi\hbar c^2 p \theta^2}.  
\label{secc.dif.LL.singular}  
\end{equation}  
They comment that the singular behavior of the cross section in  
$\theta$ 
when it goes to zero is specifically a  
quantum effect, without any further comment.  We will study this  
problem in the next sections.

%%% Section: Relativistic case %%%   
\section{Solenoidal magnetic field in the relativistic case.}  
\label{secc:rel-sol}  
  
Let us consider the scattering of a Dirac particle by the magnetic
field of a solenoid with a constant magnetic flux. This is a problem
in which free particle asymptotic states can be used. The beam
polarization will be taken into account.  As mentioned before, this is
a problem studied before by other authors also using perturbation
theory, but here our interest is quite different. We want to study
the classical limit of the result and discuss the proper way
to get it.
  
Consider a long solenoid of lenght $L$ and radius $r_0 \ll L$ centered  
in the ${\bf \hat{\i}_3}$ axis. Inside of the solenoid, where  
$r<r_0$, the magnetic field is uniform, ${\bf B}=B_0{\bf \hat{\i}_3}$,  
with $B_0$ being a constant, while outside the solenoid, where $r>r_0$, the  
magnetic field is null. For $r>r_0$, the  
magnetic flux $\Phi$ will be constant: $\Phi = \pi r^2_0 B_0$.  
We will follow the Bjorken and Drell convention~\cite{BD}.

A vector potential that describes such magnetic field for the  
interior of the solenoid is   
${\bf A_i}=-{\bf r \times B_0}/2 = rB_0{\bf\hat{\theta}}/2$  
with ${\bf\hat{\theta}}=(-\sin{\theta}, \cos{\theta}, 0)$. Outside the   
solenoid, the vector potential is  
${\bf A_o}=\Phi{\bf\nabla}\theta/2\pi$,   
with ${\bf \nabla} \times {\bf A_o}=0={\bf B}(r>r_0)$.   
Using the Levi-Civita symbol in three indices, $\epsilon_{ijk}$, the  
vector potential of the solenoid field can be written as   
$$  
\slash{A}=A_{\mu}\gamma^{\mu}=\frac{\Phi}{2\pi}\epsilon_{ij3}x_i\gamma^j  
\cases{\frac{1}{r^2_0} & for $r<r_0$ \cr  
            \frac{1}{x^2_1+x^2_2} & for $r>r_0$,}  
$$  
with the scalar potential $A^0=0$.  
Replacing this vector potential in the $S$ matrix,  
equation~(\ref{ec:Sfi.prim.ord}), for Dirac particle solutions,  
equation~(\ref{ec:psi.libre}), and $f \neq i$, we obtain  
$$  
S_{fi}=-\frac{iem\Phi}{2\pi V \sqrt{E_i E_f}} \epsilon_{ij3}  
\left[ \int_{r<r_0}{e^{iqx}\frac{x_i} {r^2_0}d^4x} +  
\int_{r>r_0}{e^{iqx}\frac{x_i} {x^2_1+x^2_2}d^4x} \right] \bar{u}_f \gamma^j u_i.  
$$  
  
Notice that there exists a global phase $e^{ie\Phi\theta/hc}$ in the  
free particle wave function because the presence of a pure gauge field  
in the exterior of the solenoid, but it does not contribute to  
$S_{fi}$. For both integrals  
the parts that correspond to $dx_0=dt$ and $dx_3$ are proportional to  
$2\pi \delta(q_0)$ and $2\pi \delta(q_3)$, respectively. So the  
energy-momentum conservation in the scattering process is guaranteed  
and the particles do not change their momentum in the magnetic field  
direction. The integrals in the plane perpendicular to ${\bf B}$ are  
\begin{eqnarray}  
\int_{r<r_0}{e^{-i{\bf q\cdot x}} x_i d^2x} &=& 2\pi i r^3_0 \frac{q_i}{q}  
   \left[ \frac{J_0(q r_0)}{q r_0} - 2\frac{J_1(q r_0)}{(q r_0)^2} \right], \nonumber \\   
\int_{r>r_0}{e^{-i{\bf q\cdot x}}\frac{x_i}{x^2_1+x^2_2}d^2x}&=& -2\pi i \frac{q_i}{q^2} J_0(q r_0), \nonumber  
\end{eqnarray}  
where ${\bf q}=q_1{\bf \hat{\i}_1} + q_2{\bf \hat{\i}_2}, {\bf  
x}=x_1{\bf \hat{\i}_1} + x_2{\bf \hat{\i}_2}$ and $J_n$  
are the $n$ order Bessel functions \cite{Arfken}. In this form, the  
$S$ matrix for $f \neq i$ is:  
$$  
S_{fi} = -\frac{2em\Phi}{V r_0 \sqrt{E_i E_f}}   
            [2\pi \delta(q_0)][2\pi \delta(q_3)]   
           J_1(q r_0) \epsilon_{ij3} \frac{q_i}{q^3}   
           \bar{u}_f \gamma^j u_i.  
$$  
We note that in the lowest order in the $S$ matrix there exists a net  
contribution from the interior of the solenoid, where the magnetic  
field is not null, in contradisctintion with the LL calculation, where  
only the vector potential of the exterior of the  
solenoid is considered.  
  
Finally, the differential cross section per unit length of the  
solenoid is  
\begin{eqnarray}  
\frac{d\sigma}{dx_3 d\theta}&=&\int{\frac{(2em\Phi)^2}{2\pi r^2_0 E_f}   
                           \delta(q_0) \delta(q_3)  
                        \frac{{\left| J_1(q r_0)\right|}^2}{p_i q^6}  
                        {\left| \bar{u}_f \slash{q} u_i \right|}^2  
                        p_f dp_f dp_{3f} }\nonumber \\  
                       &=&\frac{(2em\Phi)^2}{2\pi r^2_0}   
                          \frac{{\left| J_1(q r_0)\right|}^2}{p_i q^6}  
                          {\left| \bar{u}_f \slash{q} u_i \right|}^2.  
                          \nonumber  
\end{eqnarray}  
  
Averaging over incident polarizations we  obtain  
$$  
\frac{1}{2}\Sigma_{s_i}{\left|{ \bar{u}_f \slash{q} u_i }\right|^2} =   
        q^2(m^2-p_fp_i) + 2(p_iq)(p_fq)  
$$  
and we note that this result does not depend on the final  
polarization. After some algebra we get  
$$  
\frac{1}{2}\Sigma_{s_i}{\left|{ \bar{u}_f \slash{q} u_i }\right|^2}  =   
           16 p^4 \sin^2{\frac{\theta}{2}}  
$$  
where $p = |{\bf p_f}| = |{\bf p_i}|$. Finally, introducing $\hbar$ and $c$ explicitly, we have  
\begin{equation}   
\frac{d\sigma}{dx_3 d\theta} =   
       \frac{1}{f} \frac{\hbar}{c^2} \left({\frac{e\Phi}{r_0}}\right)^2   
       \frac{{\left| J_1(2\frac{p}{\hbar}r_0  
                     {\left|\sin{\frac{\theta}{2}}\right|})  
              \right|}^2}  
            {8\pi p^3 \sin^4{\frac{\theta}{2}}}   
\label{dsigma}   
\end{equation}   
which has the same form whether or not the final polarization of the
beam is actually measured ($f=1$ or $f=2$).  As can be observed, the
differential cross section is symmetric in $\theta$. This is
reminiscent of the Stern-Gerlach result, in which an unpolarized beam
interacting with an inhomogeneous magnetic field is equally split into
two parts, each one with opposite spin. But, as we have mentioned
before, equation~(\ref{dsigma}) does not depend on the final polarization
of the particles.  Thus, this symmetric behavior of $\theta$ should be
a consequence of the perturbation theory but notice that it is also
present in non perturbative results like those of AB and LL.
  
\begin{center}  
\begin{figure}  
\epsfig{file=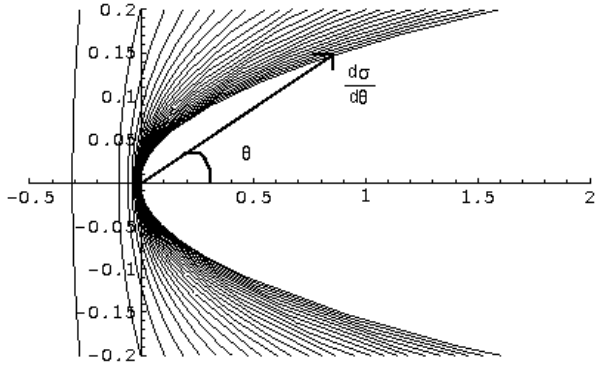,bb=85 390 295 530,clip=,angle=0,width=8cm}  
\caption{Polar plot of $\frac{d\sigma}{dx_3 d\theta} \times 10^{52}$  
         for $\Phi=4.3 \times 10^{-7}$gauss cm$^2$, one quanta of magnetic  
         flux ($n=1$), $r_0=1$cm and incident energy running from 1MeV to  
         50MeV in steps of 2MeV.}  
\label{fig-1}  
\end{figure}  
\end{center}

Figure~\ref{fig-1} shows the behavior of the cross section of
equation~(\ref{dsigma}) in a polar plot scaled by a $10^{52}$ factor for a
quanta of magnetic flux ($\Phi \sim 4\times10^{-7}$ gauss cm$^2$),
$r_0=1$cm and the energy of the incident particles running from 1~MeV
to 50~MeV in steps of 2MeV as a function of the scattering angle
$\theta$.
 
%%% Polarization %%%  
When helicity states (HS) are considered, the resulting differential cross
section is:
\begin{equation}   
\left.{\frac{d\sigma}{dx_3 d\theta}}\right|_{HS} =   
       \frac{1}{2\pi} \frac{\hbar}{c^2} \left({\frac{e\Phi}{r_0}}\right)^2   
       \frac{{\left| J_1(2\frac{p}{\hbar}r_0  
                     {\left|\sin{\frac{\theta}{2}}\right|})  
              \right|}^2}  
            {4^3 p^3 \sin^4{\frac{\theta}{2}}} (1 + \lambda_i\lambda_f)^2,
\label{dsigma_HS}   
\end{equation}   
where $\lambda_i$ and $\lambda_f$ stand for the initial and final
states of helicity and acquire only the values $+1$ or $-1$.
A zero differential cross section is obtained if
$\lambda_i = - \lambda_f$. When 
$\lambda_i = \lambda_f$, which 
implies helicity conservation, then the 
differential crosss section is:
\begin{equation}   
\left.{\frac{d\sigma}{dx_3 d\theta}}\right|_{HS} =   
       \frac{1}{4} \frac{\hbar}{c^2} \left({\frac{e\Phi}{r_0}}\right)^2   
       \frac{{\left| J_1(2\frac{p}{\hbar}r_0  
                     {\left|\sin{\frac{\theta}{2}}\right|})  
              \right|}^2}  
            {8 \pi p^3 \sin^4{\frac{\theta}{2}}},
\label{dsigma_HS(i=f)}   
\end{equation}   
which has the same form as equation~(\ref{dsigma}).

%%% Section: Non-relativistic reduction %%%  
\section{Non-relativistic reduction.}  
  
To make connection with previous results, we
study the limit case of small scattering angles. If we
assume $pr_0\left|{\sin{(\theta/2)}}\right|/\hbar\ll 1$, then
equation~(\ref{dsigma}) (or equation~(\ref{dsigma_HS(i=f)})) reduces to
\begin{equation}  
\left.{\frac{d\sigma}{dx_3 d\theta}}  
  \right|_{\frac{p}{\hbar}r_0\left|{\sin{\frac{\theta}{2}}}\right| \ll 1} =   
  \frac{1}{f}   
  \frac{e^2\Phi^2}{8\pi c^2 \hbar p \sin^2{\frac{\theta}{2}}},  
\label{secc.mm.r}  
\end{equation}  
which agrees with the result reported by Aharonov-Bohm when
$e\Phi/2\hbar c \ll 1$. And if we impose in equation~(\ref{secc.mm.r}) the
condition $\theta \ll 1$ we obtain
\begin{equation}  
\left.{\frac{d\sigma}{dx_3 d\theta}}  
  \right|_{\frac{p}{\hbar}r_0 \left|{\sin{\frac{\theta}{2}}}\right| \ll 1, 
  \theta \ll 1}   
  = \frac{1}{f} \frac{e^2\Phi^2}{2\pi c^2 \hbar p \theta^2}.  
\label{secc.mm.h.c}  
\end{equation}  
which is presicely the result reported by Landau and Lifshitz.
  
We want to point out that it does not make sense to take the Planck's  
limit ($\hbar \rightarrow 0$) in equation~(\ref{secc.mm.r}) or in  
equation~(\ref{secc.mm.h.c}), because both expressions were obtained  
assuming the condition $pr_0\left|{\sin{(\theta/2)}}\right|/\hbar \ll  
1$. Hence, we have to take the classical limit using the exprression  
for the differential cross section given in equations.~(\ref{dsigma}) 
or (\ref{dsigma_HS(i=f)}).

%%% Section: Classical limit %%%  
\section{Classical limit (Planck's limit).}  
  
Let us study now the classical limit of the differential cross section  
of equation~(\ref{dsigma}). For this purpose, consider the new adimensional  
variable $x = 2 p r_0  
{\left|\sin{(\theta/2)}\right|}/\hbar = r_0 q$ and define $J(x) =  
|J_1(x)|^2/x$. Observe that the limit $\hbar \rightarrow 0$ implies $x  
\rightarrow \infty$ or $pr_0 \rightarrow \infty$~\cite{Zkarzhinsky-97}  
for fixed 
$\theta$. We can rewrite equation~(\ref{dsigma}) as follows: 
$$  
\frac{d\sigma}{dx_3 d\theta} =   
     \frac{1}{\pi f} \left({\frac{e\Phi}{2c}}\right)^2   
     \frac{J(x)}{r_0 p^2 \left|{\sin^3{\frac{\theta}{2}}}\right|}. 
$$  
Because the asymptotic behavior of the Bessel function~\cite{Arfken} 
is:  
$$  
\lim_{x \rightarrow \infty}{J_1(x)} =   
      - \sqrt{\frac{2}{\pi x}}\cos{\left(x - \frac{3}{4}\pi\right)},   
      \hspace{.5cm}  x \gg \frac{3}{8};  
$$  
the resulting Planck's limit of equation~(\ref{dsigma})
is identically zero for fixed $e, p, r_0, \Phi$ and $\theta$:  
\begin{equation}  
\lim_{\hbar \rightarrow 0}\frac{d\sigma}{dx_3 d\theta}  
  =\lim_{\hbar \rightarrow 0}\frac{\hbar^2}{f}   
   \left({\frac{e\Phi}{2\pi c}}\right)^2  
   \frac{\cos^2{\left({2\frac{p}{\hbar}r_0  
                      {\left|\sin{\frac{\theta}{2}}\right|}   
                      - \frac{3}{4}\pi }\right)}}  
        {2 r_{0}^{3} p^4 \left|{\sin^5{\frac{\theta}{2}}}\right|}  
  =0,
\label{climit-dsigma}  
\end{equation}   
which is also obtained for $pr_0 \rightarrow \infty$.

So, the perturbative result gives a consistent finite classical limit  
and reduces to the eikonal and the zero radius limits. 
If the classical limit is attempted in
equations.~(\ref{secc.mm.r}) or (\ref{secc.mm.h.c}), the result would be
singular in $\hbar$, but this is clearly a misleading procedure.

The apparent difference in the classical limits of
equation~(\ref{dsigma}) (see equation~(\ref{climit-dsigma})) and
equations.~(\ref{secc.mm.r}) and (\ref{secc.mm.h.c}) comes from the fact
that in taking the limit $\hbar \rightarrow 0$ of the perturbative
result, equation~(\ref{dsigma}), the Bessel function decreses as $J_1(x)
\sim 1/\sqrt{x}$ and it does generate an $\hbar$ contribution to the
cross section. On the other hand, if one begins by taking
$pr_0\left|{\sin{(\theta/2)}}\right|/\hbar \ll 1$, the Bessel function  
approximates to $J_1(x) \sim x$, and this behaves like $1/\hbar$. The  
overall difference between these two procedures is an $\hbar^3$  
factor. It is important to notice that loop corrections to the  
perturbative expansion do not modify the $\hbar$ behavior of the  
amplitude, this can be proved with the use of the loop expansion.

%%% Section: Quantized magnetic flux %%%  
\section{Quantized magnetic flux.}  
  
One can obtain a non divergent expression for the Landau-Lifshitz  
and the Aharonov-Bohm results when $\hbar \rightarrow 0$ if in  
place of $e$ one fixes $\Phi$, the magnetic flux. The rationale behind  
this is that instead of a classical limit, with fixed $e$, one imposes  
the magnetic flux quantization condition, $\Phi = n\Phi_0$.  
  
For a quantized magnetic flux, where $\Phi_0 = hc/e = 4.318 \times  
10^{-7}$ gauss cm$^2$, the cross section of  
equation~(\ref{dsigma}) takes the form  
$$  
\frac{d\sigma}{dx_3 d\theta} =   
    n^2 \hbar^3 \frac{\pi}{f}    
    \frac{{\left| J_1(2\frac{p}{\hbar}r_0  
                      {\left|\sin{\frac{\theta}{2}}\right|})\right|}^2}  
         {2 r_0^2 p^3 \sin^4{\frac{\theta}{2}}},  
$$  
which apart of being independent of the charge of the particles, is a  
cross section of a purely quantum effect. Cast in this way it is not  
singular in $\hbar$ as in the form that Landau-Lifshitz report. In  
particular, for the case of small scattering angles, it takes the form  
\begin{equation}  
\left.{\frac{d\sigma}{dx_3 d\theta}}\right|_{\theta \ll 1} =   
          \frac{8 \pi^2 \hbar n^2}{f p \theta^2},  
\label{dsigmaMM.small_theta}  
\end{equation}  
which also has a null classical limit. If we recall  
equation~(\ref{secc.dif.LL.singular}) obtained by Landau-Lifshitz, we  
see that the same form can be recovered when $\Phi= n hc/e$,  
but these authors did not quantize the magnetic flux and thus  
they cannot get a cross section of a pure quantum problem as that of  
equation~(\ref{dsigmaMM.small_theta}).  
  
Also note that the zero classical limit with quantized magnetic flux  
of equation~(\ref{secc.dif.AB}) is obtained:  
$$  
\frac{d\sigma}{d\theta}=\frac{\hbar \sin^2(n\pi)}  
       {2\pi p \sin^2{\frac{\theta}{2}}}  
       \stackrel{\longrightarrow}{_{\hbar \rightarrow 0}} 0.  
$$

%%% Section: Conclusions %%%   
\section{Conclusions.}  
  
In this work we present a relativistic quantum study in first order
of perturbation theory of the cross section of the scattering of a
Dirac particle with magnetic fields. We have specially focused in the
classical limit for the solenoidal magnetic field case.
  
In order to fulfill the perturbation theory requirements the magnetic  
field was bounded to a solenoidal one with constant flux. We obtained that  
the cross section of the scattering problem is given by equation~(\ref{dsigma})  
and has the same form whether or not the final polarization of  
the beam is actually measured. This indicates that  
the symmetry in the scattering angle is most likely a consequence of the  
perturbation theory.  
  
We have shown that taking $pr_0\left|{\sin{(\theta/2)}}\right|/\hbar
\ll 1$ our result reduces to that one reported by Aharonov-Bohm.
Also, our result agrees with that one obtained by Landau-Lifshitz
when we take $pr_0\left|{\sin{(\theta/2)}}\right|/\hbar \ll 1$ and
$\theta \ll 1$. Notice that both cases lead to similar singular
behavior in $\hbar$ (see equations.~(\ref{secc.mm.r}) and
(\ref{secc.mm.h.c})).
  
We have shown that the perturbative classical limit for all scattering
angles and all radii of the solenoid with $e, \Phi, p, r_0$ and
$\theta$ fix, is identically zero, because the cross section behaves
like $\hbar^2$ and hence it is not singular in $\hbar$ as the one of
Landau-Lifshitz (see equation~(\ref{secc.dif.LL.singular})). We point out
that the same zero classical limit can be obtained in the limit $ pr_0
\rightarrow \infty$ with fixed $e, \Phi$ and $\theta$. 
Notice that the apparent difference in the classical limits comes
from the fact that the asymptotic behavior of the Bessel function
goes like $J_1(1/\hbar) \sim 1/\sqrt{1/\hbar}$ and it does generate an
$\hbar$ contribution to the cross section, while taking first the
approximation $pr_0\left|{\sin{(\theta/2)}}\right|/\hbar \ll 1$, the
Bessel function behaves like $J_1(1/\hbar) \sim 1/\hbar$. So, the
overall difference between these two procedures is an $\hbar^3$
factor.
  
When the magnetic flux is quantized, the cross section is proportional  
to $\hbar^3$ getting, again, a null classical limit. This limit can be  
recovered from the Aharonov-Bohm and Landau-Lifshitz results  
because $\Phi = n hc/e$ and then  
$$  
\sin{\left({\frac{e\Phi}{2\hbar c}}\right)} = \sin(n\pi) \equiv 0,  
$$  
an independent result of $\hbar$.

Finally we want to point out that althought our reslut is consistent
in the sense that the Aharonov-Bohm and the Landau-Lifshitz
results are recovered, there is not a direct classical correspondence
via the Planck's limit (see equation~(\ref{climit-dsigma})), because in
particular the cross section is symmetric in $\theta$. This problem is
shared also by the AB and LL solutions and is possibly solved by
higher order corrections in the {\it external} magnetic field.

%%% Acknowledgments %%%  
%RevTex: \section*{Acknowledgments}  
  
%IoP
\ack

We want to thank the helpful comments of A.~Rosado.  This work was
partially supported by CONACyT~(3097P-E), DGAPA-UNAM~(IN118600) and
DGEP-UNAM.

%%% Appendix: Uniform and constant magnetic field %%%   
\appendix
\section*{Uniform and constant magnetic field   
          in the relativistic case.}  
\setcounter{section}{1}
\label{campo.uniforme}  

Here we consider the scattering of a Dirac particle by an  
external uniform and constant magnetic field ${\bf B_0} = B_0{\bf  
\hat{\i}_3}$ with $B_0$ a constant, described by the vector potential  
$$   
{\bf A} = -x_2B_0{{\bf \hat{\i}_1}}.   
$$  
As in perturbative QCD we take initial and final states as asymptotic  
states of free particles with momentum ${\bf p}$ and spin ${\bf s}$,  
explicitly:  
\begin{equation}  
\psi({\bf x}) =   
  \sqrt{\frac{m}{EV}}u({\bf p},{\bf s})e^{-i{\bf p} \cdot {\bf x}}.  
\label{ec:psi.libre}  
\end{equation}  
At first order in perturbation theory the scattering matrix $S_{fi}$  
is given by  
\begin{equation}  
S_{fi} =   
 \delta_{fi} -   
  ie\int{\bar{\psi}_f(y)\slash{A}(y)\psi_i(y) d^4y},  
\label{ec:Sfi.prim.ord}  
\end{equation}  
where $\bar{\psi} = \psi^{\dagger}\gamma^0$ and $\slash{A} =  
A_\mu\gamma^\mu$ with $\gamma^\mu$ the $4 \times 4$ Dirac matrices. If  
the initial state has momentum $p$ in the ${{\bf \hat{\i}_1}}$  
direction, the $S$ matrix for $f \neq i$ is  
$$  
S_{fi} =   
\frac{iemB_0}{\sqrt{E_iE_f}V} \bar{u}_f \gamma^1 u_i   
\int{x_2 e^{i{\bf x}\cdot({\bf p_f}-{\bf p_i})}d^4x},  
$$  
where we have denoted $u_i = u_i({\bf p_i}, {\bf s_i})$ and similarly  
for $u_f$. The integrals can be solved immediately because three of  
them are proportional to a Dirac delta function,  
while the integral in the $x_2$ direction is equal to $2\pi i  
\delta(q_2)/q_2$. Then, the $S$ matrix of this problem is  
$$  
S_{fi} =   
- \frac{emB_0}{\sqrt{E_iE_f}V} \frac{(2\pi)^4\delta^4({\it q})}{q_2}   
\bar{u}_f \gamma^1 u_i  
$$  
where ${\it q \equiv p_f-p_i}$ is the momentum transfer.   
  
With the usual replacement of ${\left|{2\pi\delta(q_i)}\right|}^2 =  
2\pi L_i\delta(q_i)$, we obtain the differential cross section per  
unit of magnetic field volume:  
\begin{equation}  
\frac{d\sigma}{d\theta}= 2\pi (emB_0)^2   
                         \frac{\delta(q_1)\delta(q_2)}  
                              {p^3\sin^2{\theta}}  
               {\left|\bar{u}_f \gamma^1 u_i\right|}^2.  
\label{secc.dif.campo.unif}  
\end{equation}  
  
Note that the resulting differential cross section is proportional to  
a Dirac delta function of the momentum transfer in the incident  
direction of the particles, meaning that the particles do not change  
their momentum after their interaction with the magnetic field. This  
is in apparent contradiction with the common sense, because we know  
that in the classical situation a particle that interacts with a  
magnetic field changes its momentum, and its orbit is a  
circumference. Also note that we assumed that the magnetic field fills  
all the space and we used free particle solutions to solve the  
scattering problem. These conditions are physically questionable  
because the presence of the magnetic field binds the particles and  
therefore the asymptotic sates cannot be plane waves, just like is  
done in perturbative QCD.  
  
To use perturbation theory it is necessary to verify its applicability  
limits. For perturbation theory to be valid one must  
satisfy $\left|U\right| \ll \hbar^2/ma^2$~\cite{Joachain} where $U$ is  
significant in the range $a$ and $m$ is the mass of the particle. For  
the case we are studying the range of the potential is infinite, so to  
use perturbation theory we need to modify the potential in such a way  
that it goes rapidly to zero at infinity, doing the system compatible  
with our calculation and with physics, and then the particles can be  
treated as free asymptotically.  
  
Then, a more natural way to study the scattering of particles by an  
external magnetic field, is to confine the field in space, as was done  
in section~\ref{secc:rel-sol}. Curiously enought, we notice that the  
$r_0 \rightarrow \infty$ limit of equation~(\ref{dsigma}) gives precisely  
the result of equation~(\ref{secc.dif.campo.unif}).

%%% References %%%

%%% Figures %%%  
  
%\begin{center}  
%\begin{figure}  
%\epsfig{file=plot_2.ps,bb=85 390 295 530,clip=,angle=0,width=8cm}  
%\caption{Polar plot of $\frac{d\sigma}{dx_3 d\theta} \times 10^{52}$  
%         for $\Phi=4.3 \times 10^{-7}$gauss cm$^2$, one quanta of magnetic  
%         flux ($n=1$), $r_0=1$cm and incident energy running from 1MeV to  
%         50MeV in steps of 2MeV.}  
%\label{fig-1}  
%\end{figure}  
%\end{center}  
  
\end{document}